\documentclass[preprint,pre,aps,amsfonts,amsmath,showpacs,superscriptaddress,nofootinbib, draft]{revtex4}
\usepackage{amsmath,amssymb,amscd}
\usepackage{bm}
\begin{document}
\title{Linear Amplifier Breakdown and Concentration Properties of a Gaussian Field Given that its $\bm{L^2}$-Norm is Large}
\author{Philippe Mounaix}
\email{mounaix@cpht.polytechnique.fr}
\author{Pierre Collet}
\email{collet@cpht.polytechnique.fr}
\affiliation{Centre de Physique Th\'eorique, UMR 7644 du CNRS, Ecole
Polytechnique, 91128 Palaiseau Cedex, France.}
\date{\today}
\begin{abstract}
In the context of linear amplification for systems driven by the square of a Gaussian noise, we investigate the realizations of a Gaussian field in the limit where its $L^2$-norm is large. Concentration onto the eigenspace associated with the largest eigenvalue of the covariance of the field is proved. When the covariance is trace class, the concentration is in probability for the $L^2$-norm. A stronger concentration, in mean for the sup-norm, is proved for a smaller class of Gaussian fields, and an example of a field belonging to that class is given. A possible connection with Bose-Einstein condensation is briefly discussed.
\end{abstract}
\maketitle
%
\newtheorem{lemma}{Lemma}
\newtheorem{proposition}{Proposition}
\section{Introduction}\label{sec1}
This paper is devoted to the characterization of  the realizations of a Gaussian field on a finite domain of ${\mathbb R}^d$ in the limit where its $L^2$-norm is large. Our primary motivation is to get a better understanding of linear amplification in systems driven by the square of a Gaussian noise. Nevertheless, the possibility of interpreting our results in terms of Bose-Einstein condensation indicates that this study may be of interest in a much wider range of physical situations. We will briefly elaborate on this interpretation at the end of Sec.\ \ref{sec5}. In this introductory part we restrict ourselves to the random amplifier setting.

A good starting point to explain the problem we are interested in is the work by Mounaix, Collet, and Lebowitz (MCL)\ \cite{MCL}. MCL investigated the divergence of the average solution to the stochastic PDE,
\begin{equation}\label{eq1.1}
\left\lbrace\begin{array}{l}
\partial_t{\cal E}(x,t)-\frac{i}{2m}\Delta
{\cal E}(x,t)=
\lambda\vert\varphi(x,t)\vert^2{\cal E}(x,t),\\
t\ge 0,\ x\in {\mathbb T}_L^d,\ {\rm and}\ {\cal E}(x,0)=1,
\end{array}\right.
\end{equation}
where $m$ is a complex mass with ${\rm Im}(m)\ge 0$, $\lambda >0$ is the coupling constant, $\varphi$ is a zero mean complex Gaussian noise, and ${\mathbb T}_L^d$ is a $d$-dimensional torus of length $L$. Using a distributional formulation for the solution to\ (\ref{eq1.1}), MLC proved that for a finite and non zero $m$, the value of $\lambda$ at which the $q^{\rm th}$ moment of $\vert {\cal E}(x,t)\vert$ w.r.t. $\varphi$ diverges is given by $\lambda_q=1/q\, \sup_{x(\cdot)\in B(x,t)}\kappa_1\lbrack x(\cdot)\rbrack$, where $B(x,t)$ is the set of all the continuous paths in ${\mathbb T}_L^d$ arriving at $x(t)=x$, and $\kappa_1\lbrack x(\cdot)\rbrack$ is the largest eigenvalue of the covariance of $\varphi(x(\tau),\tau)$ for $0\le\tau\le t$. The question then arises whether the presence of the non local quantity $\sup_{x(\cdot)\in B(x,t)}\kappa_1\lbrack x(\cdot)\rbrack$ in the expression for $\lambda_q$ is the sign of a corresponding non local structure (i.e. large scale) in the realizations of $\varphi$ giving rise to a large amplification.

The results given in the present paper answer that question in the two particular cases $m^{-1}=0$ and $m\rightarrow 0$. The general case of a finite and non zero $m$ considered by MCL is still out of our reach. For $m^{-1}=0$, the solution to\ (\ref{eq1.1}) at fixed $x$ reads
\begin{equation}\label{eq1.2}
{\cal E}(x,t)=\exp\left(\lambda\|\varphi(x,\cdot)\|_{2,\lbrack 0,\, t\rbrack}^2\right),
\end{equation}
where $\|\cdot\|_{2,\lbrack 0,\, t\rbrack}$ denotes the $L^2$-norm on $\lbrack 0,\, t\rbrack$. The corresponding $\lambda_q$ is easily found to be given by $\lambda_q=1/q\kappa_1$, where $\kappa_1$ is the largest eigenvalue of the covariance of $\varphi(x,\tau)$ for $0\le\tau\le t$ and fixed $x$. For $m\rightarrow 0$, it is easy to show that ${\cal E}(x,t)$ reduces to
\begin{equation}\label{eq1.3}
{\cal E}(x,t)=\exp\left(\frac{\lambda}{L^d}\|\varphi\|_{2,{\mathbb T}_L^d\times\lbrack 0,\, t\rbrack}^2\right),
\end{equation}
where $\|\cdot\|_{2,{\mathbb T}_L^d\times\lbrack 0,\, t\rbrack}$ is the $L^2$-norm on ${\mathbb T}_L^d\times\lbrack 0,\, t\rbrack$. In this case, $\lambda_q$ is given by $\lambda_q=L^d/q\kappa_1$, where $\kappa_1$ is now the largest eigenvalue of the covariance of $\varphi(x,\tau)$ for $(x,\tau)\in{\mathbb T}_L^d\times\lbrack 0,\, t\rbrack$. Expressions\ (\ref{eq1.2}) and\ (\ref{eq1.3}) are formally identical, and in both cases the divergence of the moments of $\vert {\cal E}(x,t)\vert$ is determined by the realizations of $\varphi$ with an arbitrarily large $L^2$-norm on the appropriate domain. Thus, for both $m^{-1}=0$ and $m\rightarrow 0$, the problem reduces to the investigation of the realizations of a Gaussian field in the limit where its $L^2$-norm is large.

In the context of laser-plasma interaction, the question was first addressed in\ \cite{MD} heuristically and numerically for $m^{-1}=0$. It was shown there that the realizations of $\varphi$ with a large $L^2$-norm tend to have a {\it non random} profile, $\hat{\varphi}\equiv\varphi/\|\varphi\|_2$, given by the (normalized) eigenfunction associated with $\kappa_1$, assumed not to be degenerate. In the present work, we investigate the problem from a rigorous mathematical point of view and we extend the results to the case of a degenerate $\kappa_1$.

Although this paper is written in a mathematical style, it is intended to a physics-oriented audience and we do not claim that our results are particularly deep from a mathematical point of view. The important thing is not so much the mathematics as the physical implications of our conclusions. By revealing that the realizations of a Gaussian field that may cause the breakdown of a linear amplifier are {\it delocalized} modes, our results overturn conventional wisdom
\footnote{Most theoretical models dealing with stochastic amplification beyond the perturbative regime are hot spot models. In view of our results, the implicit assumption about the leading role of hot spots underlying all these models should be carefully reexamined.}
that breakdown is due to {\it localized} high-intensity peaks of the Gaussian field (the so-called high-intensity speckles, or hot spots\ \cite{RD}).

The outline of the paper is as follows. In Section\ \ref{sec2} we specify the class of $\varphi$ we consider and we give some necessary definitions. Section\ \ref{sec3} deals with the concentration of $\varphi$ in probability onto the eigenspace associated with $\kappa_1$ when $\|\varphi\|_2$ is large. A stronger concentration is established in Section\ \ref{sec4} for a smaller class of $\varphi$. Finally, the connection between our results and Bose-Einstein condensation is briefly discussed in Section\ \ref{sec5}.
%
%
\section{Definitions}\label{sec2}
Let $\varphi(x)$ be a complex Gaussian field
\footnote{This is the case of interest in laser-plasma interaction and nonlinear optics in which $\varphi$ is the (complex) time-enveloppe of the laser electric field. With the help of some minor modifications, our results carry over straightforwardly to the cases where $\varphi$ is real.}
on a bounded subset of $\mathbb{R}^d$, $\Lambda$, with zero mean, ${\rm Cov}[\varphi(x),\varphi(y)]=0$, and ${\rm Cov}[\varphi(x),\varphi(y)^\ast]=C(x,y)$. Let $T_C$ be the covariance operator acting on $f(x)\in L^2(\Lambda)$, defined by (using Dirac's bracket notation)
\begin{equation}\label{eq2.1}
\langle x\vert T_C\vert f\rangle=\int_\Lambda C(x,y)f(y)\, d^dy ,
\end{equation}
with $x,\, y\in\Lambda$. Write $\mu_1\ge\mu_2\ge\cdots >\mu_n\ge\cdots$ the eigenvalues of $T_C$, $\phi_n(x)$ the corresponding orthonormal eigenfunctions, and $\kappa_1>\kappa_2>\cdots >\kappa_n>\cdots$ the distinct values taken by the $\mu_n$ with degeneracies $g_1,\, g_2,\cdots ,g_n,\cdots$. Let $\lbrace s_n\rbrace$ be a sequence of i.i.d. complex Gaussian random variables with zero mean, ${\rm Cov}(s_n ,s_m)=0$, and ${\rm Cov}(s_n^\ast ,s_m)=\delta_{nm}$. We consider the class of $\varphi(x)$ which can be written as a Karhunen-Lo\`eve expansion\ \cite{Adl}
\begin{equation}\label{eq2.2}
\varphi(x)=\sum_{n=1}^{+\infty}s_n\sqrt{\mu_n}\phi_n(x),
\end{equation}
with $\mu_n\searrow 0$ fast enough as $n\nearrow +\infty$ (to be specified later on). From a physical point of view, condition\ (\ref{eq2.2}) is quite generic as every centered Gaussian field with a continuous $C(x,y)$ has an expansion of this form\ \cite{Adl}.

Finally, we write $d\mathbb{P}$ the Gaussian probability measure on the appropriate function space\footnote{Typically $L^2(\Lambda)$ or $C^0(\Lambda)$, depending on the speed at which $\mu_n\searrow 0$ as $n\nearrow +\infty$.} of $\varphi$ and $\mathbb{E}$ the corresponding expectation.
%
%
\section{Concentration of $\bm{\hat{\varphi}}$ in probability for a large $\bm{\|\varphi\|_2}$}\label{sec3}
In this section we characterize the structure of $\varphi(x)$ by proving concentration of the profile $\hat{\varphi}(x)\equiv\varphi(x)/\|\varphi\|_2$ in probability when $\|\varphi\|_2$ is large. Let the subscripts $\parallel$ and $\bot$ respectively denote the projections onto and orthogonal to the $g_1$-dimensional eigenspace associated with $\kappa_1$. One has,
\begin{proposition}\label{prop1}
If $T_C$ is trace class, then for every $\varepsilon >0$,
\begin{equation}\label{eq3.1}
\lim_{r\rightarrow +\infty}\mathbb{P}\left(\left.\|\hat{\varphi}_{\bot}\|_2>\varepsilon\, \right\vert\, \|\varphi\|_2^2>r \right)=0.
\end{equation}
\end{proposition}
{\it Proof.} Propositions\ \ref{prop1} and\ \ref{prop2} can be proved by using a general large deviation principle for Gaussian random functions\ \cite{Lif}. Nevertheless, since not all physicists are familiar with the machinery of large deviation principles, we give here equivalent proofs requiring no prior knowledge of this theory. For every $r>0$,
\begin{equation}\label{eq3.2}
\mathbb{P}\left(\left.\|\hat{\varphi}_{\bot}\|_2>\varepsilon\, \right\vert\, \|\varphi\|_2^2>r \right)\le
\mathbb{P}\left(\left.\frac{\|\varphi_{\bot}\|_2}{\sqrt{r}}>\varepsilon\, \right\vert\, \|\varphi\|_2^2>r \right).
\end{equation}
Let $d\mathbb{P}_{\bot}$ and $d\mathbb{P}_{\parallel}$ denote the probability measures of $\|\varphi_{\bot}\|_2^2$ and $\|\varphi_{\parallel}\|_2^2$, respectively. By statistical independence of $\varphi_{\bot}$ and $\varphi_{\parallel}$ one has,
\begin{equation}\label{eq3.3}
\mathbb{P}\left(\frac{\|\varphi_{\bot}\|_2}{\sqrt{r}}>\varepsilon\, ,\, \|\varphi\|_2^2>r \right)=
\int_{u=\varepsilon^2 r}^{+\infty}d\mathbb{P}_{\bot}(u)\int_{v=r}^{+\infty}d\mathbb{P}_{\parallel}(v-u).
\end{equation}
From\ (\ref{eq2.2}) one gets,
\begin{equation}\label{eq3.4}
d\mathbb{P}_{\parallel}(v)=\frac{H(v)v^{g_1-1}}{(g_1-1)!\kappa_1^{g_1}}\exp\left(-\frac{v}{\kappa_1}\right)\, dv,
\end{equation}
where $H(v)$ is the Heaviside step function. Since $H(v)v^{g_1-1}$ is an increasing function of $v$ it follows from\ (\ref{eq3.4}) that, for every $u\ge 0$,
\begin{equation*}
d\mathbb{P}_{\parallel}(v-u)\le\exp\left(\frac{u}{\kappa_1}\right)\, d\mathbb{P}_{\parallel}(v).
\end{equation*}
Thus, (\ref{eq3.3}) is bounded by
\begin{eqnarray}\label{eq3.5}
\mathbb{P}\left(\frac{\|\varphi_{\bot}\|_2}{\sqrt{r}}>\varepsilon\, ,\, \|\varphi\|_2^2>r \right)&\le&
\mathbb{P}\left( \|\varphi_{\parallel}\|_2^2>r \right)
\int_{u=\varepsilon^2 r}^{+\infty}\exp\left(\frac{u}{\kappa_1}\right)\, d\mathbb{P}_{\bot}(u) \nonumber \\
&\le&\mathbb{P}\left( \|\varphi\|_2^2>r \right)
\int_{u=\varepsilon^2 r}^{+\infty}\exp\left(\frac{u}{\kappa_1}\right)\, d\mathbb{P}_{\bot}(u),
\end{eqnarray}
and from\ (\ref{eq3.2}) and\ (\ref{eq3.5}) one gets
\begin{equation}\label{eq3.6}
\mathbb{P}\left(\left.\|\hat{\varphi}_{\bot}\|_2>\varepsilon\, \right\vert\, \|\varphi\|_2^2>r \right)\le
\int_{u=\varepsilon^2 r}^{+\infty}\exp\left(\frac{u}{\kappa_1}\right)\, d\mathbb{P}_{\bot}(u).
\end{equation}
It remains to prove that the right-hand side of\ (\ref{eq3.6}) tends to zero as $r\rightarrow +\infty$. By exponential Markov inequality, one has for every positive $a<1/\kappa_2$,
\begin{eqnarray}\label{eq3.7}
\mathbb{P}\left(\|\varphi_{\bot}\|_2^2>u\right)&\le&{\rm e}^{-au}
\mathbb{E}\left\lbrack\exp\left(a\|\varphi_{\bot}\|_2^2\right)\right\rbrack \nonumber \\
&=&{\rm e}^{-au}\prod_{n\ge 2}\left(1-a\kappa_n\right)^{-g_n}.
\end{eqnarray}
The existence of the product on the right-hand side of\ (\ref{eq3.7}) is insured by $T_C$ being trace class. Now, by taking $a=(\kappa_1^{-1}+\kappa_2^{-1})/2$ it can be seen from\ (\ref{eq3.7}) that $\mathbb{P}\left(\|\varphi_{\bot}\|_2^2>u\right)$ is bounded above by a constant times $\exp\lbrack -(\kappa_1^{-1}+\kappa_2^{-1})u/2\rbrack$. As a result, $\exp(u/\kappa_1)$ is $d\mathbb{P}_{\bot}$-integrable which completes the proof of Proposition\ \ref{prop1}. $\square$
%
%
\section{A stronger concentration of $\bm{\hat{\varphi}}$ for a large $\bm{\|\varphi\|_2}$}\label{sec4}
The concentration of $\hat{\varphi}(x)$ onto the fundamental eigenspace of $T_C$ can be made stronger by considering a smaller class of $\varphi(x)$. This is the subject of the following proposition, where $\|\cdot\|_{\infty}$ denotes the uniform norm on $\Lambda$.
\begin{proposition}\label{prop2}
Assume $\varphi(x)$ is a.s. continuous and $\langle x\vert T_C^{1/2}\vert x\rangle$ is bounded in $\Lambda$, then,
\begin{equation}\label{eq4.1}
\lim_{r\rightarrow +\infty}\mathbb{E}\left(\left.\|\hat{\varphi}_{\bot}\|_{\infty}\, \right\vert\, \|\varphi\|_2^2>r \right)=0.
\end{equation}
\end{proposition}
{\it Proof.} Let $\psi(x)$ be the Gaussian field defined by
\begin{equation}\label{eq4.2}
\psi(x)=\sum_{n=1}^{+\infty}s_n(\mu_n/\kappa_1)^{1/4}\phi_n(x).
\end{equation}
Since $\varphi(x)$ is a.s. continuous in $\Lambda$, the convergence in\ (\ref{eq2.2}) is pointwise with probability one\footnote{It is even uniform over $\Lambda$, with probability one. See Theorem 3.1.2. in Ref.\ \cite{Adl}.}. Thus, from\ (\ref{eq2.2}), the Schwartz inequality, and\ (\ref{eq4.2}) one gets $\vert\varphi_\bot(x)\vert\le\kappa_1^{1/4}\|\psi_\bot\|_2\langle x\vert T_C^{1/2}\vert x\rangle^{1/2}$, for every $x$ in $\Lambda$. Since $\langle x\vert T_C^{1/2}\vert x\rangle$ is assumed to be bounded in $\Lambda$, $\exists\, B>0$ such that $\sup_{x\in\Lambda}\langle x\vert T_C^{1/2}\vert x\rangle^{1/2} \le\kappa_1^{1/4}B$, (the factor $\kappa_1^{1/4}$ has been introduced for convenience), and one has
\begin{equation}\label{eq4.3}
\|\hat{\varphi}_\bot\|_{\infty}\le\sqrt{\kappa_1}B\, \|\psi_\bot\|_2/\|\varphi\|_2.
\end{equation}
It follows from\ (\ref{eq4.3}) and $\|\varphi\|_2\le\sqrt{\kappa_1}\, \|\psi\|_2$ [which follows from\ (\ref{eq2.2}) and\ (\ref{eq4.2})] that, for every $\varepsilon >0$,
\begin{equation}\label{eq4.4}
\mathbb{P}\left(\|\hat{\varphi}_{\bot}\|_{\infty}>\varepsilon\, ,\, \|\varphi\|_2^2>r \right)\le
\mathbb{P}\left(\frac{\|\psi_{\bot}\|_2}{\sqrt{r/\kappa_1}}>\frac{\varepsilon}{B}\, ,\, \|\psi\|_2^2>\frac{r}{\kappa_1} \right).
\end{equation}
From\ (\ref{eq3.3}) with $\varepsilon =0$ and\ (\ref{eq3.4}) one gets,
\begin{equation}\label{eq4.5}
\mathbb{P}\left(\|\varphi\|_2^2>r\right)=\int_{u=0}^{+\infty}\exp\left(\frac{u}{\kappa_1}\right)\, d\mathbb{P}_{\bot}(u)
\int_{v=r}^{+\infty}\frac{H(v-u)(v-u)^{g_1-1}}{(g_1-1)!\kappa_1^{g_1}}\exp\left(-\frac{v}{\kappa_1}\right)\, dv ,
\end{equation}
which yields in the large $r$ limit,
\begin{eqnarray}\label{eq4.6}
\mathbb{P}\left(\|\varphi\|_2^2>r\right)&\sim&\left(\frac{r}{\kappa_1}\right)^{g_1-1}
\frac{\exp(-r/\kappa_1)}{(g_1-1)!}
\int_{u=0}^{+\infty}\exp\left(\frac{u}{\kappa_1}\right)\, d\mathbb{P}_{\bot}(u)\ \ \ \ (r\rightarrow +\infty) \nonumber \\
&=&\left(\frac{r}{\kappa_1}\right)^{g_1-1}
\frac{\exp(-r/\kappa_1)}{(g_1-1)!}\prod_{n\ge 2}\frac{1}{(1-\kappa_n/\kappa_1)^{g_n}}.
\end{eqnarray}
One finds similarly
\begin{equation}\label{eq4.7}
\mathbb{P}\left(\|\psi\|_2^2>\frac{r}{\kappa_1}\right)\sim\left(\frac{r}{\kappa_1}\right)^{g_1-1}
\frac{\exp(-r/\kappa_1)}{(g_1-1)!}\prod_{n\ge 2}\frac{1}{(1-\sqrt{\kappa_n/\kappa_1})^{g_n}}\ \ \ \ (r\rightarrow +\infty),
\end{equation}
and
\begin{equation}\label{eq4.8}
\lim_{r\rightarrow +\infty}\frac{\mathbb{P}\left(\|\psi\|_2^2 >r/\kappa_1\right)}{\mathbb{P}\left(\|\varphi\|_2^2 >r\right)}
=\prod_{n\ge 2}\left(
\frac{1-\kappa_n/\kappa_1}{1-\sqrt{\kappa_n/\kappa_1}}\right)^{g_n}\stackrel{def}{=}C_\infty .
\end{equation}
Note that since $\langle x\vert T_C^{1/2}\vert x\rangle$ is bounded in $\Lambda$ and $\vert\Lambda\vert <+\infty$, $T_C^{1/2}$ is trace class, which insures the existence of $C_\infty$. As a result, there is a constant $C>C_\infty$ such that, for $r$ large enough,
\begin{equation}\label{eq4.9}
\mathbb{P}\left(\|\psi\|_2^2>\frac{r}{\kappa_1}\right)\le C\mathbb{P}\left(\|\varphi\|_2^2>r\right),
\end{equation}
and by\ (\ref{eq4.4}),\ (\ref{eq4.9}), and the proof of Proposition\ \ref{prop1} for $\psi$,
\begin{equation}\label{eq4.10}
\lim_{r\rightarrow +\infty}
\mathbb{P}\left(\left.\|\hat{\varphi}_{\bot}\|_{\infty}>\varepsilon\, \right\vert\, \|\varphi\|_2^2>r \right)=0.
\end{equation}
From the obvious inequality $\|\hat{\varphi}_{\bot}\|_{\infty}\le\varepsilon +\|\hat{\varphi}_{\bot}\|_{\infty}\bm{1}_{\lbrace\|\hat{\varphi}_{\bot}\|_{\infty} >\varepsilon\rbrace}$ and\ (\ref{eq4.3}) one gets the estimate $\|\hat{\varphi}_{\bot}\|_{\infty}\le\varepsilon +\sqrt{\kappa_1}B\, \|\psi_\bot\|_2\|\varphi\|_2^{-1} \bm{1}_{\lbrace\|\hat{\varphi}_{\bot}\|_{\infty} >\varepsilon\rbrace}$, which gives, after conditional averaging,
\begin{eqnarray}\label{eq4.11}
\mathbb{E}\left(\left.\|\hat{\varphi}_{\bot}\|_{\infty}\, \right\vert\, \|\varphi\|_2^2>r \right)&\le&
\varepsilon +B\mathbb{E}\left(\left.\sqrt{\kappa_1}
\frac{\|\psi_\bot\|_2}{\|\varphi\|_2} \bm{1}_{\lbrace\|\hat{\varphi}_{\bot}\|_{\infty} >\varepsilon\rbrace}
\, \right\vert\, \|\varphi\|_2^2>r \right) \nonumber \\
&\le&\varepsilon +B\mathbb{E}\left(\left.
\frac{\|\psi_\bot\|_2}{\sqrt{r/\kappa_1}} \bm{1}_{\lbrace\|\hat{\varphi}_{\bot}\|_{\infty} >\varepsilon\rbrace}
\, \right\vert\, \|\varphi\|_2^2>r \right) \\
&\le&\varepsilon +B\mathbb{E}\left(\left.
\frac{\|\psi_\bot\|_2^2}{r/\kappa_1}\, \right\vert\, \|\varphi\|_2^2>r \right)^{1/2}
\mathbb{P}\left(\left.\|\hat{\varphi}_{\bot}\|_{\infty}>\varepsilon\, \right\vert\, \|\varphi\|_2^2>r \right)^{1/2}. \nonumber
\end{eqnarray}
Using $\|\psi_\bot\|_2\le\|\psi\|_2$ and $\|\varphi\|_2\le\sqrt{\kappa_1}\|\psi\|_2$ one has,
\begin{equation}\label{eq4.12}
\mathbb{E}\left(\left.\frac{\|\psi_\bot\|_2^2}{r/\kappa_1}\, \right\vert\, \|\varphi\|_2^2>r \right)\le
\frac{\kappa_1}{r\mathbb{P}(\lbrace\|\varphi\|_2^2>r)}
\int_{r/\kappa_1}^{+\infty} x\mathbb{E}\lbrack\delta(\|\psi\|_2^2 -x)\rbrack\, dx.
\end{equation}
Since $\mathbb{P}(\|\psi\|_2^2>r/\kappa_1)$ behaves like $\exp(-r/\kappa_1)$ as $r\rightarrow +\infty$ [to within algebraic corrections, see\ (\ref{eq4.7})], the integral on the right-hand side of\ (\ref{eq4.12}) behaves like
\begin{equation}\label{eq4.13}
\int_{r/\kappa_1}^{+\infty} x\mathbb{E}\lbrack\delta(\|\psi\|_2^2 -x)\rbrack\, dx\sim\frac{r}{\kappa_1}
\mathbb{P}\left(\|\psi\|_2^2>\frac{r}{\kappa_1}\right)\ \ \ \ (r\rightarrow +\infty).
\end{equation}
Now,\ (\ref{eq4.12}),\ (\ref{eq4.13}), and\ (\ref{eq4.8}) yields
\begin{equation}\label{eq4.14}
\limsup_{r\rightarrow +\infty}\mathbb{E}\left(\left.\frac{\|\psi_\bot\|_2^2}{r/\kappa_1}\, \right\vert\, \|\varphi\|_2^2>r \right)
\le C_\infty .
\end{equation}
Then, from\ (\ref{eq4.11}),\ (\ref{eq4.10}), and\ (\ref{eq4.14}) one gets
\begin{equation}\label{eq4.15}
\limsup_{r\rightarrow +\infty}\mathbb{E}\left(\left.\|\hat{\varphi}_{\bot}\|_{\infty}\, \right\vert\, \|\varphi\|_2^2>r \right)\le\varepsilon .
\end{equation}
It remains to take $\varepsilon$ arbitrarily small and\ (\ref{eq4.15}) reduces to\ (\ref{eq4.1}), which completes the proof of Proposition\ \ref{prop2}. $\square$

The following proposition provides an example of a class of $\varphi$ for which Proposition\ \ref{prop2} holds. Take $d=1$ and $\Lambda =\lbrack 0,\, 1\rbrack$. One has,
\begin{proposition}\label{prop3}
If $C(x,y)$ has the fourth order partial derivative $(\partial^4/\partial x^4)C(x,y)$ continuous on $\lbrack 0,\, 1\rbrack^2$, then Proposition\ \ref{prop2} holds.
\end{proposition}
{\it Proof.} First we prove that $\langle x\vert T_C^{1/2}\vert x\rangle$ is bounded in $\Lambda$. From the inequalities $\|\phi_n\|_\infty^2\le\|\phi_n\|_2^2+2\|\phi_n\|_2\|\phi_n^\prime\|_2$, $\|\phi_n^{(p)}\|_2\le\|\phi_n^{(2p)}\|_2^{1/2}\|\phi_n\|_2^{1/2}$, and the normalization $\|\phi_n\|_2=1$ one gets $\|\phi_n\|_\infty^2\le 1+2\|\phi_n^{(4)}\|_2^{1/4}$, which yields the estimate
\begin{equation}\label{eq4.16}
\langle x\vert T_C^{1/2}\vert x\rangle =\sum_{n=1}^\infty \sqrt{\mu_n}\, \vert\phi_n(x)\vert^2
\le\sum_{n=1}^\infty \sqrt{\mu_n}\, (1+2\|\phi_n^{(4)}\|_2^{1/4}).
\end{equation}
Since $(\partial^4/\partial x^4)C(x,y)$ is continuous on $\lbrack 0,\, 1\rbrack^2$ it is also bounded on $\lbrack 0,\, 1\rbrack^2$ and there is a positive constant $a$ (independent of $n$) such that
\begin{eqnarray*}
\|\phi_n^{(4)}\|_2 &=&\frac{1}{\mu_n}\left\lbrack\int_0^1\left\vert\int_0^1
\frac{\partial^4 C(x,y)}{\partial x^4}\phi_n(y)\, dy\right\vert^2 dx\right\rbrack^{1/2} \\
&\le&\frac{1}{\mu_n}\left\lbrack\int_0^1\int_0^1\left\vert
\frac{\partial^4 C(x,y)}{\partial x^4}\right\vert^2 dx\, dy\right\rbrack^{1/2}\le\frac{a}{\mu_n}.
\end{eqnarray*}
Injecting this inequality into the right-hand side of\ (\ref{eq4.16}) and using the fact that under the conditions of Proposition\ \ref{prop3}, $\mu_n=o(1/n^5)$ as $n\rightarrow +\infty$\ \cite{CH}, one finds
\begin{equation}\label{eq4.17}
\langle x\vert T_C^{1/2}\vert x\rangle
\le\sum_{n=1}^\infty \left(\mu_n^{1/2}+2a^{1/4}\mu_n^{1/4}\right)<+\infty .
\end{equation}
We now prove that $\varphi(x)$ is a.s. continuous. Since $C(x,y)$ is continuous (and bounded) on $\lbrack 0,\, 1\rbrack^2$, it follows from
\begin{equation*}
\vert\phi_n(x)-\phi_n(x^\prime)\vert\le\frac{1}{\mu_n}
\left\lbrack\int_0^1\vert C(x,y)-C(x^\prime ,y)\vert^2 dy\right\rbrack^{1/2},
\end{equation*}
and dominated convergence that $\phi_n(x)$ is continuous on $\lbrack 0,\, 1\rbrack$. Thus, for any given integer $N>0$, $\sum_{n=1}^{N}s_n\sqrt{\mu_n}\phi_n(x)$ is a.s. continuous on $\lbrack 0,\, 1\rbrack$. Now, by Borel-Cantelli lemma and the asymptotic behavior of $\mu_n$ for large $n$ (see above), one has $\vert s_n\vert\le 1/\mu_n^{1/8}$ a.s. as $n\rightarrow +\infty$. This result, together with the inequalities $\|\phi_n\|_\infty\le 1+\sqrt{2}\, \|\phi_n^{(4)}\|_2^{1/8}$ and $\|\phi_n^{(4)}\|_2\le a/\mu_n$, yield $\vert s_n\vert\sqrt{\mu_n}\, \vert\phi_n(x)\vert\le\mu_n^{3/8}+\sqrt{2}\, a^{1/8}\mu_n^{1/4}$ a.s. as $n\rightarrow +\infty$. Therefore,\ (\ref{eq2.2}) converges uniformly over $\lbrack 0,\, 1\rbrack$ with probability one, and $\varphi(x)$ is a.s. continuous. $\square$
%
%
\section{Summary and perspectives}\label{sec5}
In this paper, we have studied the realizations of a Gaussian field, $\varphi$, in the limit where its $L^2$-norm is large. We have first proved concentration of $\varphi$ in probability onto the eigenspace associated with the largest eigenvalue of the covariance of $\varphi$, $\kappa_1$, when $\|\varphi\|_2$ is large (Proposition\ \ref{prop1}). Considering then a smaller class of $\varphi$, we have established a stronger type of concentration (Proposition\ \ref{prop2}). Finally, we have given an example of a class of $\varphi$ for which that stronger type of concentration holds (Proposition\ \ref{prop3}). These results extend the heuristic and numeric results of\ \cite{MD} and give them a mathematically rigorous meaning.

We can now answer the question we asked in Sec.\ \ref{sec1} for $m^{-1}=0$ and $m\rightarrow 0$. The reason for the presence of $\kappa_1$ in the expression for $\lambda_q$ when ${\cal E}(x,t)$ is given by\ (\ref{eq1.2}) or\ (\ref{eq1.3}) is clear: the divergence of the moments of $\vert{\cal E}(x,t)\vert$ is determined by realizations of $\varphi$ that concentrate onto the eigenspace associated with $\kappa_1$. Those realizations do have a large scale stucture encoded in the reduction of the spectrum of $\varphi$ as it concentrates onto the $\kappa_1$-eigenspace. The components orthogonal to the $\kappa_1$-eigenspace do not play any role in the onset of the divergence of the moments of $\vert{\cal E}(x,t)\vert$.

A possible physical interpretation of our results follows from the similarity between our problem and the spherical model of a ferromagnet\ \cite{BK}. The role of the spins in the spherical model is played by the $s_n$ in\ (\ref{eq2.2}), and the fixed magnetization constraint is replaced with a fixed $\|\varphi\|_2^2$. The well known connection between the spherical model and the ideal Bose gas\ \cite{GB} then suggests that the concentration of $\varphi$ for a large $\|\varphi\|_2^2$ may be interpreted as a Bose-Einstein condensation in which the eigenspace associated with $\kappa_1$ plays the role of the ground state in the Bose gas. The following point seems to plead in favor of this interpretation. Define,
\begin{equation}\label{eq5.1}
\begin{array}{l}
\mathcal{E}_{\vert\vert}(r)=\mathbb{E}\left\lbrack\left.\|\varphi_\parallel\|_2^2\, \right\vert\, \|\varphi\|_2^2>r \right\rbrack ,\\
\mathcal{E}_{\bot}(r)=\mathbb{E}\left\lbrack\left.\|\varphi_\bot\|_2^2\, \right\vert\, \|\varphi\|_2^2>r \right\rbrack .
\end{array}
\end{equation}
Using the estimate [see\ (\ref{eq3.6})],
\begin{equation*}
\mathbb{P}\left(\left.\|\varphi_{\bot}\|_2^2>v\, \right\vert\, \|\varphi\|_2^2>r \right)\le
\int_{u=v}^{+\infty}\exp\left(\frac{u}{\kappa_1}\right)\, d\mathbb{P}_{\bot}(u),
\end{equation*}
one has,
\begin{eqnarray*}
\mathcal{E}_{\bot}(r)&=&\int_0^{+\infty}v\, d\mathbb{P}_{\bot}\left(\left. v\, \right\vert\, \|\varphi\|_2^2>r \right) \\
&=&\int_0^{+\infty}\mathbb{P}\left(\left.\|\varphi_{\bot}\|_2^2>v\, \right\vert\, \|\varphi\|_2^2>r \right)\, dv \\
&\le&\int_0^{+\infty}dv\, \int_{u=v}^{+\infty}\exp\left(\frac{u}{\kappa_1}\right)\, d\mathbb{P}_{\bot}(u) \\
&=&\int_0^{+\infty}u\, \exp\left(\frac{u}{\kappa_1}\right)\, d\mathbb{P}_{\bot}(u)<+\infty ,
\end{eqnarray*}
where we have used the fact that $d\mathbb{P}_{\bot}$ is absolutely continuous with a density behaving like $\exp(-r/\kappa_2)$ as $r\rightarrow +\infty$ [to within algebraic corrections, see\ (\ref{eq4.6}) in which $\kappa_i$ and $g_i$ are respectively replaced with $\kappa_{i+1}$ and $g_{i+1}$]. Since this estimate is independent of $r$, $\rho_\bot =\sup_{r\in\mathbb{R}^+}\mathcal{E}_{\bot}(r)<+\infty$, and from the inequality $\mathcal{E}_{\vert\vert}(r)+\mathcal{E}_{\bot}(r)>r$ [see\ (\ref{eq5.1})] one obtains,
\begin{equation}\label{eq5.2}
\begin{array}{l}
\mathcal{E}_{\vert\vert}(r)>r-\rho_\bot , \\
\mathcal{E}_{\bot}(r)\le\rho_\bot .
\end{array}
\end{equation}
It follows immediately from\ (\ref{eq5.2}) that for $r\gtrsim 2\rho_\bot$, the only contribution of the eigenspace associated with $\kappa_1$ is greater than the one of all the other eigenmodes (which remains bounded). Such a behavior is typical of a Bose-Einstein condensation onto the eigenspace associated with $\kappa_1$.

Of course, like any other phase transition, no sharp condensation can occur in a finite size system. The ``thermodynamic" limit, $\vert\Lambda\vert\rightarrow +\infty$ with fixed $\|\varphi\|_2^2/\vert\Lambda\vert$, must be taken to get unambiguous results. In this limit, one is faced with the problem that the $\kappa_n$, with $n\ge 2$, get closer and closer to $\kappa_1$ as $\vert\Lambda\vert\rightarrow +\infty$, which makes it difficult to tell them apart from $\kappa_1$ and may jeopardize condensation by leading to a concentration onto a larger space than the eigenspace associated with $\kappa_1$. For a homogeneous field\footnote{i.e. with correlation function $C(x,y)=C(x-y)$.}, one expects that issue to be all the more acute as the density of states at large wavelengths, close to the ground state (here, the condensate), is large. This will be the case at low space dimensionality $d$. Such a dimensional effect is well-known in traditional Bose-Einstein condensation of an ideal Bose gas which needs $d\ge 3$ to exist. A thorough study of the concentration properties of $\varphi$ in the thermodynamic limit, from the Bose-Einstein condensation point of view, is currently in progress along the same line as Evans {\it et al.}\ \cite{EMZ}. This will be the subject of a future work.
\section*{Acknowledgements}
Ph. M. thanks Satya N. Majumdar and Alain Comtet for providing valuable insights, and in particular for pointing out the connection with Bose-Einstein condensation. The authors also thank Joel L. Lebowitz for useful discussions on related subjects.
%
%

%
%
\end{document}